\documentclass[11pt]{article}
\pdfoutput=1
 \usepackage{textcomp}  
\usepackage{amsmath,amssymb} 
\usepackage{epsfig,latexsym}
    \usepackage{graphicx}
\newcommand{\sect}[1]{\setcounter{equation}{0}\section{#1}}
\renewcommand{\theequation}{\arabic{section}.\arabic{equation}}
\def\am{angular-momentum~}
\def\amm{angular-momenta~}
\def\ab{\al\bt}
\def\al {\alpha}
\def\avv{angular velocity~}
\def\avvv{angular velocities~}

\def\AAA{{\cal A}}  
 
\def\ba{\begin{eqnarray}}
\def\bal{\mbox{\bm$\al$\ubm}}
 \def\bB{\mbox{\bm$B$\ubm}}
 \def\bbb{background~}
\def\bbbb{backgrounds~}
\def\be{\begin{equation}}
\def\bep{\mbox{\bm$\ep$\ubm}}
\def\bga{\mbox{\bm$\ga$\ubm}}
\def\bGa{\mbox{\bm$\Ga$\ubm}}
 \def\bg{{\bf g}}
 \def\bla{\mbox{\bm$\la$\ubm}}
\def\bLa{\mbox{\bm$\La$\ubm}}
\def\bi{\bibitem}
\def\bm{\boldmath}
\def\bde{\mbox{\bm$\de$\ubm}~}
\def\bJ{\mbox{\bm$J$\ubm}~}
\def\bOm{\mbox{\bm$\Om$\ubm}~}
\def\bom{\mbox{\bm$\om$\ubm}~}
\def\bte{\mbox{\bm$\te$\ubm}}
\def\bxi{\mbox{\bm$\xi$\ubm}}
\def\bt {\beta}
\def\br{{\bf r}} 
\def\bU{{\mbox{\bm$U$\ubm}~}}
\def\bu{\bullet}
\def\bV{{\mbox{\bm$V$\ubm}~}}
\def\Bar {\overline}
\def\BB{{\cal B}} 
\def\BD{\Bar D}
\def\Bg{\Bar g} 
\def\Bna{\Bar\na}
\def\BGa{\Bar\Ga}
\def\BR{\Bar R}
\def\ccc{cosmological~}
\def\cd{\cdot}
\def\cl{\centerline}
\def\co{coordinate~}
\def\coo{coordinates~}
\def\CC{{\cal C}}
\def\de{\delta}

\def\dga{\dot{g}}
\def\di{\partial}
\def\tLL{\dot{\LL}}
\def\tna{ \dot{\na}}
\def\tRR{\dot{\RR}}
\def\tw{\td w}
\def\tV{\dot{V}}
\def\VV{{\cal V}}

\def\DD{{\cal D}}
\def\De{\Delta}

\def\ea{\end{eqnarray}} 
\def\ee{\end{equation}}
\def\eee{equation~}
\def\eeee{equations~}
\def\em{energy-momentum~}
\def\en{energy~}
\def\ep{\epsilon} 
\def\eq{\equiv~}
\def\et{\eta}

\def \EE{{\cal E}}

\def\fr{\frac}

\def\FF{{\cal F}}

 \def\ga{\gamma}
 \def\gi{gauge invariant~}
 \def\gmn{g_{\mu\nu}}
 \def\gre{gravitational energy~}

\def\Ga{\Gamma}
\def\gaa{\dot{\Ga}}
\def\GR{General Relativity~}
\def\ha{\frac{1}{2}~}
 \def\hxi{\hat \xi}
 \def\hg{\hat g}
\def\hJ{\hat J}
\def\hG{\hat G}
\def\hk{\hat k}
\def\hLa{\hat\La}
\def\hLL{\hat\LL}
\def\hJ{\hat J}
\def\hJJ{\hat\JJ}
\def\hU{\hat U}
\def\hV{\hat V}
\def\hW{\hat W}   
 \def\hy{hypersurface~}
\def\HH{{\cal H}}

\def\iii{inertial frame~}
\def\iiii{inertial frames~}
\def\in{\infty}

\def\II{{\cal I}}

\def\JJ{{\cal J}}

\def\ka{\kappa}
\def\kk{{\cal K}}
\def\kkk{Killing field~}
\def\kkkk{Killing fields~}

\def\la {\lambda}
\def\lb{\label} 
\def\lhs{left-hand side~}
\def\li{\limits}
\def\lll{\left(}

\def\L{$\bullet\triangleright$~}
\def\La {\Lambda}
\def\LL{{\cal L}}
\def\LLL{\left[}

\def\mb{\mbox}
\def\me{mechanical energy~}
\def\mn{\mu\nu}

\def\MM{{\cal M}}

\def\na{\nabla}
\def\nBB{\not\!\BB}
\def\ndi{\not\!\di}
\def\nGa{\not\!\Ga}
\def\nl{\newline}
\def\nn{\nonumber}  
\def\nna{\not\!\nabla}
\def\nnn{\noindent}
\def\nEE{\not\!\EE}
\def\nHH{\not\!\!\HH}
\def\np{\newpage}
\def\nR{\not\!\! R}

\def\NN{{\cal N}}

\def\om{\omega}
\def\oo{\over}

\def\Om {\Omega}
\def\OO {{\cal O}}

\def\po{\pounds}
\def\pp{\cal P }
\def\ppp{perturbation~}
\def\pppp{perturbations~}

\def\ra{\rightarrow}
\def\rhs{right-hand side~}
\def\rh{\rho}
\def\rr{{\cal R }}
\def\rrr{\right)}
\def\rs{\rho\si}

\def\R{$~\triangleleft\bullet$~}
\def\RR{{\cal R}}
\def\Ra{\Rightarrow}
\def\RRR {\right]}

\def\si{\sigma}
\def\sq{\sqrt}
\def\sqg{\sqrt{-g}}
\def\sqBg{\sqrt{\Bg}} 
\def\sim{\simeq}
\def\sss{spacetime~}
\def\ssss{spacetimes~}
\def\st{stationary~}
\def\sup{superpotential~}
\def\supp{superpotentials~}

 \def\Si{\Sigma}
\def\Sc{Schwarzschild~}
\def\Sup{Superpotential~}

\def\td{\tilde}
\def\tDe{\tilde\La} 
\def\tEE{\tilde\EE}
\def\tga{\tilde\ga}
\def\tGa{\tilde\Ga}
\def\te{\theta}
\def\tf{\tilde f}
\def\th {\frac{1}{3}}
\def\tha{\ts{\ha}}
\def\ti{\times}
\def\tk{\td k}
\def\tLL{\tilde\LL}
\def\tna{\tilde\na}
\def\ts{\textstyle}
\def\tRR{\tilde\RR}
\def\tU{\tilde U}
\def\tV{\td V}
\def\tW{\td W}
\def\tw{\td w}

\def\TT{{\cal T}}
\def\Te{\Theta}
\def\ubm{\unboldmath}
\def\un{\underline}
\def\unb{\underbrace}
\def\us{\underset}
\def\UU{{\cal U}}

\def\vAAA{\vec{\AAA}}
\def\vf{\varphi }
\def\vCC{\vec\CC}
\def\vEE{\vec{\EE}}
\def\vHH{\vec{\HH}}
\def\vJ{{\vec J}}
\def\vom{\vec \omega}
\def\vOm{\vec \Omega}
\def\vp{\varpi}
\def\vr{{\vec r}}
\def\vna{\vec{\na}}
\def\vs{\vskip 0.5 cm}

\def\we{\wedge}
\def\wrt{with respect to~}
\def\WW{{\cal W}}

\def\XX{{\cal X}}

\def\YY{{\cal Y}}

\def\ze{\zeta}

\def\ZZ{{\cal Z}}

\def\1k{\fr{1}{\ka}}

\def\2k{\fr{1}{2\ka}}

\def\4k{\fr{1}{4\ka}}

\title{{\bf Affine Gravity, Palatini Formalism and Charges\footnote{The work is  dedicated to Joshua Goldberg from whom I   learned and got interested in conservation laws in General Relativity (J.K).}  }}
\author{  Joseph Katz$^1$\thanks{email:jkatz@phys.huji.ac.il
}  \,and Gideon I.  Livshits$^2$\thanks{email: livshits.gideon@mail.huji.ac.il} \\
 \\  {\it  $^1$The Racah Institute of Physics}
 \\
 \\
{\it $^2$ Institute of Chemistry }
\\
\\
 {\it The Hebrew University, Givat Ram, 91904 Jerusalem, Israel}}
 
\begin{document}
\maketitle
\begin{abstract} 
  %________________ABSTRACT____________

\setlength{\baselineskip}{20pt plus2pt}

\vs
Affine gravity and the Palatini formalism contribute both to produce a simple and unique formula for calculating charges at spatial and null infinity for Lovelock type Lagrangians whose variational derivatives do not depend on second-order derivatives of the field components. The method is based on the covariant generalization due to Julia and Silva of the Regge-Teitelboim procedure that was used to define properly the mass in the classical formulation of Einstein's theory of gravity. Numerous applications reproduce standard results obtained by other secure but mostly specialized methods. As a novel application we calculate the Bondi energy loss in five dimensional gravity, based on the asymptotic solution given by Tanabe, Tanahashi and Shiromizu, and obtain, as expected, the same result. We also give the \sup for Einstein-Gauss-Bonnet gravity and find the superpotential for Lovelock theories of gravity when the number of dimensions tends to infinity with  maximally symmetrical boundaries. 

\nnn The paper is written in  standard component formalism. 
\vs
\nnn Keywords: affine gravity, Palatini formalism, Lovelock Gravity, charges at null infinity.

\end{abstract}

%----------------------------- Section 1 is Introduction -----------------------------
\setlength{\baselineskip}{20pt plus2pt}
\sect{Introduction}
% 1 (i) Hamiltonian formalism and the conservation of charges-------------------------

{\it (i) Hamiltonian formalism and the conservation of charges}
\vs
The way to evaluate total energy  in the Hamiltonian formulation of general relativity \cite{ADM}  was put on a sound basis by Regge and Teitelboim \cite{RT} for localized sources\footnote{Localized matter or fields that decrease faster than $1/r$ at flat spatial infinity.} in spacetimes that are asymptotically flat. The correct Hamiltonian $H$ has the following particularities: (1) The field equations are derived from a volume integral generator. (2) The generator   is  zero as a consequence of the  equations, i.e. on-shell. (3) For any solution of Einstein's \eeee with given boundary conditions, $H$ reduces to  a surface integral on the boundary and with the boundary   at spatial infinity the Hamiltonian is equal to  the total mass-energy of the sources $E$ including gravitational energy\footnote{ The method has been used in a variety of contexts, for a relatively recent application of the Regge-Teitelboim method see, for instance,   Jamsin \cite{Ja}.}.

In the  covariant Hamiltonian formalism or covariant phase space as described by Julia and Silva\footnote{See also   \cite{JS00} which  contains most    references to original works on the subject.} \cite{JS02} the Hamiltonian appears in a slightly different light: (1) $H$ is now  associated with a timelike global vector field $\bxi$, and is an integral over a spacelike hypersurface of a   vector density\footnote{The density character is indicated by a hat ``~\^~" over a symbol.} $\bf \hJ$. (2) The generator of the field equations is an integral over a spacelike hypersurface  of a vector density current $\bf \hW$, a linear combination of variational derivatives which is thus   equal to zero on-shell. (3)  The Hamiltonian on-shell reduces to a surface integral  on the boundary $S$ of the domain of integration $V$. The surface  integral is generated  by an anti-symmetric tensor density  $\bf \hU$  defined on the boundary; this is the superpotential. The surface integral is   equal to the energy $E$ provided the timelike vector field becomes pure time translations at spatial flat infinity.

We have thus the following sequence of equalities\footnote{Greek indices go from $0$ to $N-1$, $N$ is the dimension of \sss with signature $-(N-2)$. $N\ge 4$.} in covariant phase space:
\ba
&&  \hJ^\mu=\hat W^\mu+\di_\nu\hU^{\mn}~~~\Ra~~~ H=\int_V\hJ^\mu dV_\mu=\int_V   \hW^\mu dV_\mu+\oint_S \hU^{\mn} dS_{\mn};   \nn
\\
&&{\rm on~ shell}~~~\hW^\mu=0~~~,~~~\di_\mu\hJ^\mu=0;~~~{\rm and}~~~ H=E=\oint_S \hU^{\mn} dS_{\mn}.
\lb{11}
\ea
 
 In both cases $E$  is the time component of a 4-vector in Minkowski spacetime\footnote{Various examples of applications can be found in \cite{JS00}.}.
 If the timelike vector field $\bxi$ is replaced by   a spacelike vector field which turns (fast enough) into space translations along one of the spatial axes at infinity, the surface integral will be a component of the linear momentum along that axis. The way to generate angular momentum components is now obvious. 
 
 The   formalism holds for fields other than gravity with other global or asymptotic symmetries which are   associated with other constants of motion or conserved quantities or what are   generally called charges\footnote{The word charge without qualification may  have different meaning for different authors. Here we used the word charge as it is understood, for instance,  in Hollands, Ishibashi and Marolf \cite{HIM} as constants of motions,  or non constant if at null infinity, associated with asymptotic (and global) symmetries. Iyer and Wald \cite{IW} define     Noether charges. These are not quite the same and are more like some generalizations of Komar's covariant charges.}. The whole machinery works also on a   hypersurface   at null infinity and allows one to calculate time-dependent  quantities of physical relevance like the Bondi mass \cite{Bo},   the Sachs linear momentum \cite{Sa}  and the  \am \cite{KL97} with its usual ambiguities in four dimensions but apparently with no ambiguity in 5 dimensions \cite{TTS11}.

 %_________(ii) The superpotential_______________________
 \vs
 {\it (ii) The superpotential }
 \vs
 The superpotential $ \bf \hU$ generates thus the conserved or non-conserved charges associated with the symmetries or Killing vectors of the metric on the boundary.

There are two problems with the \sup$\!\!$. 

(1) It must be calculated in Minkowski coordinates\footnote{See for instance Goldberg's \cite{Go} calculation of the Bondi Mass.} in spite of the fact that spherical coordinates are more convenient on a sphere, but in other coordinates the surface integral most often diverges.  To avoid the latter  one needs to introduce a background \cite{Ka}, \cite{KBL} or counter-terms \cite{HIM}, \cite{MMV} or use a regularization procedure \cite{KO} or  some     other device \cite{ACOTZ}. 

(2) More importantly, the \sup is not unique because the Lagrangian is not unique. 
 
We shall remove the coordinate difficulty by introducing a background metric $\bf \Bg$. The utility  of backgrounds was suggested by Rosen \cite{Ro} in the 1940's and often used since then in various publications, but   we shall not take it as far as Rosen himself. The background metric $\bf \Bg$ is defined as follows:  take the metric $\bf g$   to be a solution of Einstein's \eeee that satisfies the boundary conditions  and remove the source terms from the solution. Make them equal to zero. The result  must be the metric far from the sources and thus on  or near the boundary (at spatial infinity) ${\bf g}=\bf\Bg$. Notice that the definition  of the  \bbb  is perfectly covariant. We shall never use the \bbb as a ``background" metric anywhere else but on the boundary\footnote{In some cases, in particular in cosmological \ssss$\!\!$, a background turns out to be useful \cite{AS}.}, so we shall not have to define a mapping of our \sss onto the background. If \sss  is asymptotically flat, $\bf \Bg$  is the Minkowski metric in any appropriate coordinates used for the solution of Einstein's equations. We then define   the  superpotential with respect to that background: ${\bf \hU(g)}-{\bf \hU(\Bg)}$. This is  a covariant tensor which reduces to $\bf \hU(g)$ in Minkowski coordinates because in those coordinates ${\bf \hU(\Bg)}=0$. The background solves the coordinate problem   and allows us to calculate the total energy $E$  in arbitrary coordinates at infinity, most usually in spherical coordinates. 
 
 The superpotential non-uniqueness due to the non-uniqueness of the Lagrangian density  is a different problem. It is best resolved by the method suggested by Regge and Teitelboim\footnote{See also Henneaux and Teitelboim \cite{HT} for anti-de Sitter backgrounds.} \cite{RT}.   The {\it variation} of the Hamiltonian should have no boundary terms by analogy with classical mechanics.   Thus the Hamiltonian $H$ should contain a boundary term $\oint \bf \hU d\bf S $ but its variation $\de H$ should not. This will define $\de\bf{\hU(g)}$. Since the effective Hamiltonian generator is a linear combination of variational derivatives, $\de\bf{\hU(g)}$ thus defined is independent of any divergence added to the Lagrangian. Hence the superpotential is   defined on the boundary up to an integration  ``constant". 
 A complete unique \sup is then obtained by taking the integration ``constant"   equal to   $-{\bf \hU(\Bg)}$.
  %--------------------(iii) Superpotentials, Affine Gravity and Palatini formalism --------------------------
 \vs 
 {\it (iii) Superpotentials, Affine Gravity and the Palatini formalism }
 \vs
   Silva \cite{Si}   generalized the Regge-Teitelboim idea about the superpotential  to the covariant Hamiltonian formulation of a large class of field theories, those in which the Lagrangians  {\it and} their variational derivatives depend at most on first-order derivatives  of the field components; moreover,  the Lie derivatives  of the fields with respect to a vector field  $\bxi$ depend at most on first-order derivatives of $\bxi$ as well.  Therefore, application to metric theories of gravity  was done in affine-$GL(N;{\rm \!I\!R})$ gravity. The more familiar and practical  Palatini formalism  would not do because the Lie derivatives of the connection coefficients, the $\bGa$'s, contain second order-derivatives of $\bxi$.  
   
   The Silva covariant generalization of the Regge-Teitelboim method as well as subsequent developments \cite{JS02} and applications were presented in terms of forms,   within the ``covariant symplectic" formulation. Our first intention here is to develop the formalism in component notations. 
   
   Affine gravity introduces an extra set of $N$ vectors\footnote{Latin indices which enumerate the vectors vary also from $0$ to $N-1$.}  $\bte^a$ which constitute a local reference frame at each point of spacetime. This frame is arbitrary and the Lagrangian is invariant under any linear transformations $\bte^a(x)=\bLa^a_{~b}(x)\bte^b(x)$. One motivation for this detour was to ensure that all Lie derivatives depend at most on first-order derivatives of $\bxi$. Affine gravity is however not   necessary   to obtain the same results \cite{KL08}. Therefore in a second step we shall review how to get the \sup directly in the Palatini formalism and in component notations, in affine gravity and in the Palatini formulation.
   
   The formula for the \sup we arrive at uses both affine gravity and Palatini formalism and as far as we are aware of is new. We also give various applications, some of which are new.
    
    %--------------------(iv) Comment ---------------------------
    \vs
{\it (iv) Comment}
\vs

We want to emphasize that the \sup exists only  on the boundaries, which may be at spatial infinity or at null infinity. The closed boundary may be anywhere but in practice it should always be taken far away from the sources of gravity otherwise the value of the Hamiltonian is meaningless because gravitational energy is  not localizable.

 %_____________________ Sect 2 - Elements -------------------------------------------
\sect{ Lagrangian variations in Palatini variables and in    affine gravity}

     In the Palatini formulation of gravity theories the fields are the metric of \sss $\bf g$ with components $\gmn$ and the   symmetric connection\footnote{Non symmetric connections are often considered also.} $\bf \Ga$ with components $\Ga^\la_{~\mn}=\Ga^\la_{~\nu\mu}$; these fields are independent.   The   curvature tensor    does not have all the symmetries of the Riemann tensor, in particular   $\di_\mu \Ga_{\nu\la}^\la\ne\di_\nu \Ga_{\mu\la}^\la$. To avoid any confusion we shall denote the curvature tensor, like in Eisenhart \cite{Ei}, by $B^\la_{~\nu\rs}$ and keep the usual notation of the Riemann tensor $R^\la_{~\nu\rs}$ for when the connection coefficients are Christoffel symbols. $B^\la_{~\nu\rs}$  is antisymmetrical in the last two indices only, $B^\la_{~\nu\rs}= - B^\la_{~\nu\si\rho}$. Additional symmetry properties, similar to those of the Riemann curvature tensor, are:
 \be
 B^\la_{~(\nu\rs)}=0~,~B^\la_{~\nu(\rs;\mu)}=0~~{\rm and}~~D_\la B^\la_{~\nu\rs}=D_{\rh} B_{\nu\si} - D_{\si} B_{\nu\rh} ~~{\rm where}~~B_{\si\nu}=B^\la_{~\si\la\nu}\ne B_{\nu\si}.
 \lb{21}
 \ee
An occasional semi-column represents a covariant differentiation. The curvature tensor itself $B^\la_{~\nu\rs}$ is given by
\be
B^\la_{~\nu\rs}= 2(\di_{[\rho}\Ga^\la_{~\si]\nu} + \Ga^\la_{~\mu[\rho}\Ga^\mu_{~\si]\nu}). 
\lb{22}
\ee
Now consider a  Lagrangian     scalar density $\hLL$ which depends on the metric components and the connection coefficients
\be
 \hLL=\sq{-g}\LL(g^{\mn}, B^\la_{~\nu\rs}).
 \lb{23}
 \ee
  $\LL$ is a scalar and   a hat over a symbol means it is multiplied by  $\sq{-g}=\sq{ - \det( g_{\mn}) }$. The matter Lagrangian for isolated systems is not necessary in what follows. This implies that we take the boundary of \sss outside of the matter, in practice at spatial or null infinity, or that fields tend to infinity faster than $1/r$.
  
  Field \eeee are the variational derivatives equated to zero. The variation of $\hLL$ for arbitrary variations of the field components $\de g^{\mn}$ and $\de \Ga^\la_{~\mn}$ may thus be written in the following form
\be
\de\hLL=*\hLL_{\mn}\de g^{\mn}+\hLL_\la^{~\mn}\de\Ga^\la_{~\mn}+\di_\mu\hat\EE^\mu.
\lb{24}
\ee
The content of $\hat\EE^\mu$ depends on the Lagrangian density which is defined  up to a divergence and will not concern us here. The variational derivatives are
\be
*\hLL_{\mn}=\fr{\di\hLL}{\di g^{\mn}}~~~{\rm and}~~~\hLL_\la^{~\mn}=\fr{\de \hLL}{\de\Ga^\la_{~\mn}};
\lb{25}
\ee
$\di$ stands for partial derivatives and $\de$ for variational derivatives.

In affine metric-theories, the central element of the description is a set of $N^2$ vector field components $\te^a_{~\mu}$ associated with a local basis in the  local tangent spaces.  There are $N(N+1)/2$ scalar products denoted by $\ga^{ab}$:
\be
\ga^{ab}=g^{\mn}\te^a_{~\mu}\te^b_{~\nu}.
\lb{26}
\ee
 The symmetric connection  $\bf \Ga$ is replaced by a spin connection $\bom$ which is a vector:
 \be
 \om^a_{~\mu b}=\te^a_{~\la}(\di_\mu\te_b^{~\la} +\Ga^\la_{~\mn}\te^{~\nu}_b)=\te^a_{~\nu} D_\mu\te_b^{~\nu},
 \lb{27}
 \ee
 $\te^{~\nu}_b$ is the inverse matrix of $\te^a_{~\nu}$. Reciprocally,
 \be
 \Ga^\la_{~\mn}=\te^b_{~\mu}(\te^{~\la}_a\om^a_{~\nu b}-\di_\nu\te_b^{~\la}).
 \lb{28}
 \ee
 Since $\bGa$ is symmetrical in $\ts{\mn}$, the anti-symmetric part of the right hand side is of course equal to zero.
 In terms of $\bte$ and $\bom\!\!$, the curvature tensor becomes
 \be
 B^\la_{~\nu\rs}= 2\te_a^{~\la}\te^b_{~\nu}(\di_{[\rho}\om^a_{~\si]b} + \om^a_{~[\rho c}\om^c_{~\si]b})=\te_a^{~\la}\te^b_{~\nu} B^a_{~b\rs}.
 \lb{29}
 \ee
 $B^a_{~b\rs}$  depends only on $\bom\!$ and its    first-order derivatives.
 
 The Lagrangian density $\hLL$, a functional of $\bg$ and $\bGa$,  see  (\ref{22}) and (\ref{23}),  may be expressed in terms  $\bte^a, \bga^{ab}$ and $\bom^{\!\!a}_{~\!b}$. However, it will   make things somewhat simpler\footnote{Though not in the spirit of affine gravity theory!}, up to a point, in what follows if instead of $\bte^a, \bga^{ab}$ and $\bom^{\!a}_{~b}$ we use $\bte^a, \bf g$ and $\bom\!^{\!a\!}_{~b}$.    It is thus understood that 
  \be
 \hLL=\hLL(g^{\mn}, \Ga^\la_{~\mn}) =\hLL( g^{\mn}, \te^a_{~\mu}, \om^a_{~\mu b}).
 \lb{210}
 \ee
 The variations of the Lagrangian   will be written as   usual   in terms of variational derivatives which provide the field equations. Thus 
 \be
 \de\hLL=*\hLL_{\mn}\de g^{\mn}+\bu\hLL_a^{~\mu}\de\te^a_{~\mu}+\hLL_a^{~\mu b}\de\om^a_{~\mu b}+\di_\mu\hat \DD^\mu,
 \lb{211}
 \ee
 in which 
 \be
 *\hLL_{\mn}=\fr{\di\hLL}{\di g^{\mn}}~~~,~~~\bu\hLL_a^{~\mu}=\fr{\di\hLL}{\di \te^a_{~\mu}}~~~,~~~\hLL_a^{~\mu b}=\fr{\de\hLL}{\de \om_{~\mu b}^a}.
 \lb{212}
 \ee
The explicit form of $\hat\DD^\mu$,   like that of $\hat\EE^\mu$, is not important.

We shall now go back from (\ref{211}) to (\ref{24}). To this effect we first replace $\de\om^a_{~\mu b}$ in  (\ref{211})  by its variations deduced from (\ref{27}) in terms of $\de\bte$ and $\de\bf \Ga$:
   \be
\de\om^a_{~\mu b}=\de\te^a_{~\nu} D_\mu\te^{~\nu}_b+\te^a_{~\nu} D_\mu\de\te^{~\nu}_b+\te^a_{~\la}  \te^{~\nu}_b\de\Ga^\la_{\mn}.
 \lb{213}
 \ee
The third term on the right hand side of (\ref{211}) becomes, with minor obvious derivations by part in the square brackets\footnote{Partial derivatives equal   covariant derivatives  of scalars,  vector densities and anti-symmetric tensor densities.}:
\ba
&&\hLL_a^{~\mu b}\de\om^a_{~\mu b}=\lll\hLL_a^{~\mu b}D_\mu\te^{~\nu}_b \rrr\de\te^a_{~\nu}+\LLL  \di_\mu \!\!\lll \te^a_{~\nu}  \hLL_a^{~\mu b} \de\te_b^{~\nu}\rrr \! - \!D_\mu  \te^a_{~\nu}  \hLL_a^{~\mu b}\de\te^{~\nu}_b -   \te^a_{~\nu}  \di_\mu  \hLL_a^{~\mu b}\de\te^{~\nu}_b        \RRR\nn
\\
&& ~~~~~~~~~~~~~~~~+\te^a_{~\la} \hLL_a^{~\mu b} \te^{~\nu}_b\de\Ga^\la_{\mn}.
\lb{214}
\ea

This formula will take a nicer form if we introduce the following self explanatory new notations:
\be
\te^a_{~\la}  \hLL_a^{~\mu b}=\hLL_\la^{~\mu b}; ~~~{\rm similarly}~~~\hLL_\la^{~\mu b}\te_b^{~\nu}=\hLL_\la^{~\mn}~~~ ,~~~\te^a_{~\la}\!\!\bu\!\hLL_a^{~\mu}=\bu\hLL_{\la}^{~\mu}~~{\rm and}~~g^{\mn}\!\!*\hLL_{\nu\la}=*\hLL_\la^\mu.
\lb{215}
\ee
With these new notations  we may rewrite (\ref{214}) as follows
\be
\hLL_a^{~\mu b}\de\om^a_{~\mu b}=D_\nu\hLL_\la^{~\mn}\te_a^{~\la}\de\te^a_{~\mu} + \hLL_\la^{~\mn}\de\Ga^\la_{~\mn} + \di_\nu\lll  \hLL_\la^{~\mn}   \te^a _{~\mu}\de\te_a^{~\la}   \rrr\!.
\lb{216}
\ee 
Inserting (\ref{216}) back into (\ref{211}), we find that
\be
 \de\hLL=*\hLL_{\mn}\de g^{\mn}+ \hLL_\la^{~\mn}\de\Ga^\la_{~\mn}+\lll   \bu\hLL_a^{~\mu} +   \te_a^{~\la}  D_\nu\hLL_\la^{~\mn} \rrr\de\te^a_{~\mu}+\di_\mu\hat\EE^\mu.
 \lb{217}
\ee
(\ref{217}) looks like (\ref{24}) except that in the Palatini formulation, there is no $\bte$-dependence\footnote{ Suppose we take variations of the fields that are all zero on the boundary and integrate over spacetime. The volume  integral should not contain any $\de\bte$ terms since $\de\bte$ is arbitrary.}. Therefore, the factor of $\de\te^a_{~\mu}$ must be identically zero:
\be
 \bu\hLL_a^{~\mu} +  \te_a^{~\la}  D_\nu\hLL_\la^{~\mn} =0~~~{\rm or} ~~~\bu\hLL_\la^{~\mu} +   D_\nu\hLL_\la^{~\mn}  =0.
 \lb{218}
\ee
This identity will be  valuable later on: it connects a variational derivative in affine gravity with the divergence of a variational derivative in the Palatini formalism.    
Equation (\ref{217}) reduces thus to 
\be
 \de\hLL=*\hLL_{\mn}\de g^{\mn}+ \hLL_\la^{~\mn}\de\Ga^\la_{~\mn}+\di_\mu\hat\EE^\mu,
 \lb{219}
\ee
which is the same as (\ref{24}).
We notice that $*\hLL_{\mn}$ and $\hLL_\la^{~\mn}=\te_{~\la}^b\te_a^{~\mu}\hLL^{~a\nu}_b$ depend only on $\bg$ and $\bB$.
 Affine gravity and  the Palatini formalism of Lagrangians that depend only on the metric and on the curvature tensor but not on their derivatives lead to {\it equivalent} equations as   expected. The variational derivatives are {\it identical} when the vector fields $\bte$ are in the directions of the local axis and $\te^a_{~\mu}=\de^a_\mu$.   
 %_____________________ Sect 3 - symmetries and conserved current in affine gravity--------------------
\sect{Symmetries and conserved currents in affine gravity  }
% 3 (i)
\vs
{\it (i) Diffeomorphism invariance in affine gravity}
\vs
Conservation laws related to the arbitrariness of the coordinates are normally derived following Noether's procedure. One considers the Lie derivative  of the Lagrangian density $\de_L\hLL$   with respect to an arbitrary vector field $\bxi$. The Lie derivative of a scalar density is a divergence $\de_L\hLL=\di_\mu(\hLL\xi^\mu)$.    In (\ref{211}) we replace the variations of the field components $\de g^{\mn}, \de\te^a_{~\mu}$ and $\de\om^a_{~\mu b}$ by their Lie derivatives as well. These do not contain higher than first-order derivatives of $\bxi$:
\ba
&&\de_ Lg^{\mn}=(D_\la g^{\mn})\xi^ \la - 2g^{\la(\mu} D_ \la\xi^{\nu)}~~~,~~~\de_L {\te_{~\mu}^a} =(D_ \la\te^a_{~\mu})\,\xi^ \la + \te^a_ {~\la} D_\mu\xi^ \la,\nn
\\
&&~~~~~~~~~~~~~~~~~~~~~~~~~~\de_L\om^a_{{~\mu} b}=(D_ \la\om^a_{{~\mu} b})\,\xi^ \la +\om^a_{{~\la} b}D_{\mu}\xi^ \la.
\lb{31}
\ea
We may thus rewrite (\ref{211}) with Lie derivatives as follows:
 \be
  *\hLL_{\mn}\de_L g^{\mn}+\bu\hLL_a^{~\mu}\de_L\te^a_{~\mu}+\hLL_a^{~\mu b}\de_L\om^a_{~\mu b}=\di_\nu\lll\! \hLL \xi^\nu - \hat\DD^\nu\!\rrr\!.
 \lb{32}
 \ee
From (\ref{31}) it is easily seen that the divergence is of a vector density linear in $\bxi$ and $D\bxi$, say of the form
\be
 \hLL \xi^\nu  - \hat\DD^\nu=*\hJ^\nu_\la\xi^\la + *\hU^{\nu\mu}_\la D_\mu\xi^\la.
\lb{33}
\ee
The \lhs is equally linear in $\bxi$ and $D\bxi$. Thus (\ref{32}) may be rewritten as follows:
\be
\hLa_\la\xi^\la+\hLa^\mu_{~\la}D_\mu\xi^\la=\di_\nu\lll    *\hJ^\nu_\la\xi^\la + *\hU^{\nu\mu}_\la D_\mu\xi^\la    \rrr \!,   
\lb{34}
\ee
in which
\be
 \hLa_\la= *\hLL_{\mn}D_\la g^{\mn}+\bu\hLL^{~\mu}_a D_\la\te^a_{~\mu}+\hLL^{~\mu b}_a D_\la\om^a_{~\mu b}~~,~~
 \hLa^\mu_{~\la}= -2\!*\!\hLL^{\mu}_\la+\bu\hLL^{~\mu}_\la+\hLL_a^{~\mu b}\om^a_{~\la b}.
 \lb{35}
\ee
It is   interesting to notice and easy  to verify     that $ - 2\!*\!\hLL^{\mu}_\la + \bu\hLL^{~\mu}_\la$ is, in affine gravity,  the variational derivative with respect to $\te^a_{~\mu}$ times $\te^a_{~\la}$:
\be
\hLL^{~\mu}_\la= - 2\!*\!\hLL^{\mu}_\la+\bu\hLL^{~\mu}_\la=\fr{\di\hLL}{\di \te^a_{~\mu}}\te^a_{~\la}\Bigg{|}_{\bga, \bom}.
\lb{36}
\ee
Now   in (\ref{34}) replace\footnote{Like in Julia and Silva \cite{JS98}. I like this clever trick (JK)!} $\bxi(x)$ by $\bxi(x)\ep(x)$; this can always be done because $\bxi$ is arbitrary.  Then expand in terms of derivatives of $\ep(x)$.  The \lhs of (\ref{34}) looks then like this:
\be
(\hLa_\la\xi^\la+\hLa^\mu_{~\la}D_\mu\xi^\la)\ep +( \hLa^\mu_{~\la}\xi^\la )D_\mu\ep.
\lb{37}
\ee
On the other hand, the right hand side is now of the following {\it form}:
\be
\di_\nu(*\hJ^\nu\ep+*\hU^{\nu\mu}\di_\mu\ep)=(\di_\nu\!*\!\hJ^\nu)\ep+(*\hJ^\mu+\di_\nu\!*\!\hU^{\nu\mu})\di_\mu\ep+(*\hU^{\nu\mu})D_{\mn}\ep.
\lb{38}
\ee
Thus, since (\ref{37}) and (\ref{38}) are equal
\be
(\hLa_\la\xi^\la+\hLa^\mu_{~\la}D_\mu\xi^\la)\ep +( \hLa^\mu_{~\la}\xi^\la )D_\mu\ep =(\di_\nu\!*\!\hJ^\nu)\ep+(*\hJ^\mu+\di_\nu\!*\!\hU^{\nu\mu})\di_\mu\ep+(*\hU^{\nu\mu})D_{\mn}\ep.
\lb{39}
\ee
  $\ep$ is arbitrary and we may identify the factors of $\ep, \di_\mu\ep$ and of $D_{\mn}\ep$ of both sides. There follows a set of ``cascade   identities" \cite{JS98}. The factors of $\ep$ reproduce of course identity (\ref{34}). The factor of $D_{\mn}\ep$ does not appear on the \lhs; thus $*\hU^{\nu\mu}$ must be antisymmetric:
\be
*\hU^{\nu\mu}= - *\hU^{\mn}.
\lb{310}
\ee
Taking this into account, the identification of the factors of $\di_\mu\ep$ leads to the following identity:
\be
*\hJ^\mu= *\hW^\mu +\di_\nu\!*\!\hU^{\mn}~~~{\rm in~ which}~~~*\!\hW^\mu\eq\hLa^{\mu}_{~\la}\xi^\la= (\hLL^{~\mu}_\la+\hLL_a^{~\mu b}\om^a_{~\la b})\xi^\la.
\lb{311}
\ee
 The vector $*\hJ^\mu$ is conserved, i.e. divergenceless, on-shell\footnote{With sources present the \eeee are of course different,except on the boundary.} ($ \hLL^{~\mu}_\la=\hLL_a^{~\mu b} =0$). The value of the integral of $*\hJ^\mu$ on-shell  over a hypersurface $V$ is equal to the integral on the boundary $S$ of $V$ of $*\hU^{\mn}$. However the ``generator" $*\hW^\mu$ is not manifestly covariant; it depends on the choice of $\bom\!\!$. In particular in the Palatini formalism  $\te^a_{~\mu}=\de^a_\mu$ and $\te^\la_a\te^b_\nu\om^a_{~\mu b}=\Ga^\la_{~\mn}$. Thus $*\hJ^\mu$, which has much of the properties of a generator of conserved charges, is unfortunately not covariant. The way around this problem which we now describe is given in \cite{JS02}.
%----------------------3 (ii)-----------------------
 \vs
{\it (ii) Freedom of choice of the local frame}
\vs
In affine gravity theories the $\bte^a$'s are $N$ arbitrary vectors which play the role of local tangent frames but the theories are invariant for   arbitrary linear changes of basis $\bte^a(x) \ra \bLa^a_{~b}(x)\bte^b(x)$. Therefore we can play the same game with this invariance as we played with the diffeomorphism invariance.

Let $\bla^a_{~b}$ characterize an infinitesimal transformation denoted by $\de_\La$. Thus,
\be
\de_\La\te^a_{~\mu}=\la^a_{~b}\te^b_{~\mu} ~~~{\rm implying~ that}~~~\de_\La\om^a_{{~\mu} b}= - \di_\mu\la^a_{~b} - \om^a_{{~\mu} c}\la^c_{~b} + \om^c_{{~\mu} b}\la^a_{~c}.
\lb{312}
\ee
The Lagrangian density and the metric components are invariant under such transformations:
\be
\de_\La\hLL=0 ~~~,~~~\de_\La g^{\mn}=0.
\lb{313}
\ee
Thus if $\de$ is replaced by $\de_\La$ in (\ref{211}), we obtain, after some rearrangement of factors,
\be
(\hLL^{~\mu}_a\te^b_{~\mu} - \hLL^{~\mu b}_c\om^c_{~\mu a}+ \hLL^{~\mu c}_a\om^b_{~\mu c})\la^a_{~b} -\hLL^{~\mu b}_a\di_\mu\la^a_{~b} = \di_\nu(\!*\!*\!\hJ^{\nu b}_a\la^a_{~b}+\!*\!*\!\hU^{\nu\mu b}_a\di_\mu\la^a_{~b}),
\lb{314}
\ee
The $\hat\DD^\mu$ in (\ref{211})    is linear in $\la^a_{~b}$ and its derivative and therefore  the \rhs in  (\ref{314}) is  of the {\it form} written above.

We now play   the same game as before by replacing $\la^a_{~b}(x)$ with $\la^a_{~b}(x)\ep(x)$ and expanding in terms of $\ep, \di_\mu\ep$ and $D_{\mn}\ep$,   identifying the corresponding factors on both sides. The factors of $\ep$ reproduce of course  (\ref{314}). The factor of $D_{\mn}\ep$ tells us that 
\be
\!*\!*\!\hU^{\nu\mu b}_a\la^a_{~b}\eq \!*\!*\hU^{\nu\mu}= - \!*\!*\hU^{\mn}.
\lb{315}
\ee
The equality of the factors of $\di_\mu\ep$ leads to an equality similar to (\ref{311}):
\be
\!*\!*\!\hJ^{\mu b}_a\la^a_{~b}\eq \!*\!*\!\hJ^{\mu}=  -\hLL^{~\mu b}_a\la^a_{~b} +\di_\nu(\!*\!*\hU^{\mn}).
\lb{316}
\ee
Here is another non-covariant divergenceless vector on-shell  $*\!*\!\hJ^\mu$ with the properties of $*\!\hJ^\mu$, see (\ref{11}).  It is divergenceless but not covariant. 
If we now sum $*\bJ$ of (\ref{311}) and $*\!*\!\bJ$ just written, we obtain a new conserved current on shell with the same properties, which can be written as follows: 
\be
\hJ^\mu=*\hJ^\mu+*\!*\!\hJ^\mu=\hLL_\la^{~\mu}\xi^\la+\hLL_a^{~\mu b} \lll \om^a_{~\nu b}\xi^\nu - \la^a_b   \rrr +\di_\nu \hU^{\mn}~~~{\rm with}~~~\hU^{\mn}=*\hU^{\mn}+*\!*\!\hU^{\mn}.
\lb{317}
\ee
% --------------------------3 (iii)-------------------------
\vs
{\it (iii) A covariant conserved current}
 \vs
 The final step in \cite{JS00} entails the  breaking of the local frame arbitrariness. The $\bte^a$'s are kept fixed like they would in a Palatini formulation\footnote{Or, if one prefers, in an Einstein-Cartan formalism ($\te^a_{~\mu}=e^a_\mu$). }, by taking  
 \be
 \de_L\te^a_{~\mu}+\de_\La\te^a_{~\mu}=0~~, ~~\di_\mu\te^a_{~\la}=0~~{\rm and}~~\te^a_{~\mu}=\de^a_\mu~~\Ra~~~\la^a_{~b}= - \te^a_{~\la}\te_b^{~\mu}\di_\mu\xi^\la ~~,~~\te_a^{~\la}\te^b_{~\nu}\om^a_{{~\mu} b}=\Ga^\la_{~\mn}.
 \lb{318}
 \ee
 Substituting the last two equalities   in the second term of (\ref{317}) on the \rhs gives:
 \be
 \hLL_a^{~\mu b}\om^a_{~\la b}\xi^ \la +\hLL^{~\mn}_ \la\di_\nu\xi^ \la =\hLL_\eta^{~\mn}\Ga^\eta_{\la\nu}\xi^ \la +\hLL^{~\mu\la}_\nu\di_\la\xi^\nu=\hLL_ \la ^{~\mn}D_\nu\xi^ \la.
 \lb{319}
 \ee
 So, $\hat\bJ$ is now of the form
 \be
 \hJ^\mu=\hW^\mu+\di_\nu\hU^{\mn}~~{\rm with}~~\hW^\mu= \hLL_\la^{~\mu}\xi^\la+\hLL_\la^{~\mn}D_\nu\xi^\la.
 \lb{320}
 \ee

 The covariant $\hJ^\mu$ is conserved on-shell ($\hW^\mu=0$) and   the charge 
 \be
 Q=\oint_S \hU^{\mn}dS_{\mn}.
\lb{321}
\ee
Q is, of course, not yet well defined because the \sup is not well-defined. We come back to this problem with an answer below. But we shall first  show that the same current is obtained by applying Noether's prescription to the same Lagrangian density in the Palatini formalism with $\bf g$ and $\bf \Ga$ as independent fields without introducing an additional     set of $N$ vectors $\bte$. We shall follow \cite{KL08} in this matter.
%---------------------------------4 Symmetries in Palatini-------------------------
 \sect{Symmetries and conserved currents in   Palatini formalism  }
Conservation laws related to diffeomorphism invariance are derived in the same way as in affine gravity. The difference lies in an extra derivative of $\bxi$. The Lie derivative  $\de_L\hLL=\di_\mu(\hLL \xi^\mu)$ as before. The Lie derivative of $g^{\mn}$ is the same as in (\ref{31}). However the Lie derivative of the $\bGa$'s contains second order derivatives and is a tensor:
\be
\de_L\Ga^\la_{~\mn}= - R^\la_{~(\mn)\eta}\xi^\eta + D_{(\mn)}\xi^\la.
\lb{41}
\ee
The result is that instead of an identity like (\ref{34}) there is one additional term on both sides in $D_{\mn}\xi^\la$. Thus after replacing $\de$ by $\de_L$ in (\ref{24}), using (\ref{31}) and (\ref{41}), we obtain an identity of the following form:
\be
\hLa_\la\xi^\la+\hLa^{~\mu}_\la D_\mu\xi^\la+\hLL_\la^{~\mn}D_{\mn}\xi^\la=\di_\nu\!\lll\hJ^\nu_\la\xi^\la + \bu\hU^{~\nu\mu}_\la D_\mu\xi^\la + \hV_\la^{~\nu\rs}D_{(\rs)}\xi^\la\rrr 
\lb{42}   
\ee
in which
\be
\hLa_\la=*\hLL _{\mn}D_\la g^{\mn}  - \hLL_\eta^{~\mn}R^\eta_{~\mn \la}~~,~~\hLa^{~\mu}_\la= - 2*\!\hLL^{~\mu}_\la.
\lb{43}
\ee
Now in (\ref{42}) we replace $\bxi(x)$ by $\bxi(x)\ep(x)$ and, as before, we expand in terms of $\ep, \di_\mu\ep, D_{\mn}\ep$ and $D_{\la\mn}\ep$.
On the \lhs we have
\be
(\hLa_\la\xi^\la+\hLa^{~\mu}_\la D_\mu\xi^\la+\hLL_\la^{~\mn}D_{\mn}\xi^\la)\ep+(\hLa_\la^{~\mu}\xi^\la+2\hLL_\la^{~\mn}D_\nu\xi^\la)D_\mu\ep+(\hLL_\la^{~\mn}\xi^\la) D_{\mn}\ep.
\lb{44}
\ee
On the right hand side we have however this:
\ba
&&\di_\nu( \hJ^\nu\ep+ \bu\hU^{\nu\mu}\di_\mu\ep+  \hV^{\nu\la\mu}D_{\la\mu}\ep)=\nn
\\
&&~~~~~~~~~~~~(\di_\nu\hJ^\nu)\ep+(\hJ^\mu+\di_\nu\bu\!\hU^{\nu\mu})\di_\mu\ep+(\bu\hU^{\nu\mu}+D_\la\hV^{\la\mn})D_{\mn}\ep+\hV^{\la\mn}D_{\la\mn}\ep.
\lb{45}
\ea
The factors of $\ep$ and its derivatives on the \lhs (\ref{44}) are identical to the factors of those derivatives on the \rhs$\!\!$ (\ref{45}). The identity  ``\lhs (\ref{44})$\eq$\!\!\rhs(\ref{45})" holds  in arbitrary coordinates. We may take \coo in which at any particular point the metric is Minkowski and the $\bGa$'s are equal to zero. At that point the last term in (\ref{45}) becomes $\hV^{\la\mn}\di_{\la\mn}\ep$. It must be zero because there is no such term on the \lhs$\!$ (\ref{44}). This implies that $\hV^{\la\mn}$,   totally symmetrized in all its indices, equals zero or that
\be
\hV^{(\la\mn)}=\ts{\fr{1}{3}}(\hV^{\la\mn}+\hV^{\mn\la}+\hV^{\nu\la\mu})=0 ~~~{\rm and, ~ see~ (\ref{42}),}~~~\hV^{\la\mn}=\hV^{\la\nu\mu}.
\lb{46}
\ee
Any $\bV\!\!$ that has these properties has also the following property which is easily proven:
\be
\hV^{\la\mn}D_{\la\mn}\ep= - \ts{\fr{2}{3}}\hV^{[\mn]\la}R^\eta_{~\la\mn}\di_\eta\ep.
\lb{47}
\ee
With this equality (\ref{47}),  we may rewrite the \rhs in (\ref{45}) as follows
\be
(\di_\nu\hJ^\nu)\ep + (\hJ^\mu+\di_\nu\!\bu\!\hU^{\nu\mu} - \ts{\fr{2}{3}}\hV^{[\rs]\nu}R^\mu_{~\nu\rs} )\di_\mu\ep+(\hLL_\la^{~\mn})D_{\mn}\ep.
\lb{48}
\ee
The \lhs  (\ref{44}) is thus also identical   to the \rhs  in the form    (\ref{48}) for any values of $\ep$: the factors of $\ep$ and its derivatives in (\ref{44})   are identical to the corresponding factors in (\ref{48}). 

The factors of $\ep$ are obtained by setting $\ep=1$ which  reproduces thus  exactly (\ref{42}) from which we learn nothing new. The two remaining identities obtained by identifying the factors of   $\di_\mu\ep$ and $D_{\mn}\ep$, are respectively
\ba
&&- 2*\!\hLL^\mu_\la\xi^\la+2\hLL_\la^{~\mn}D_\nu\xi^\la=\hJ^\mu+D_\nu\!\bu\!\hU^{\nu\mu}  - \ts{\fr{2}{3}}\hV^{[\rs]\nu}R^\mu_{~\nu\rs},
\lb{49}
\\
&& \hLL_\la^{~\mn}\xi^\la = \bu\hU^{(\mn)}+D_\la\hV^{\la\mn}.
\lb{410}
\ea
We shall decompose $\bu\hU^{\nu\mu}$ appearing  in (\ref{49}) into a symmetric   $\bu\hU^{(\nu\mu)}$ and an anti-symmetric part $\bu\hU^{[\nu\mu]}$ in $\ts{\mn}$. The symmetric part will be replaced by its value given in (\ref{410}), i.e.
\be
 \bu\hU^{(\mn)}= \hLL_\la^{~\mn}\xi^\la - D_\la\hV^{\la\mn}. 
 \lb{411}
\ee
The \rhs of (\ref{49}) then becomes
\be
\hJ^\mu+\di_\nu \bu\!\hU^{[\nu\mu]}  + \hLL_\la^{~\mn}\xi^\la - D_\la\hV^{\la\mn}- \ts{\fr{2}{3}}\hV^{[\rs]\nu}R^\mu_{~\nu\rs}.
\lb{412}
\ee
However, the last two terms combine to give 
\be
- D_\la\hV^{\la\mn}- \ts{\fr{2}{3}}\hV^{[\rs]\nu}R^\mu_{~\nu\rs}=\ts{\fr{2}{3}}\di_\nu\lll   D_\la\hV^{[\mn]\la}       \rrr\!,
\lb{413}
\ee
so that the \rhs of  (\ref{49}), i.e. (\ref{412}), is now of the form
\be
\hJ^\mu - \di_\nu\hU^{\mn}  + \hLL_\la^{~\mn}\xi^\la ~~~{\rm in~ which}~~~\hU^{\mn}= - \hU^{\nu\mu}=\bu\hU^{[\mn]} - \ts{\fr{2}{3}}  D_\la\hV^{[\mn]\la}.
\lb{414} 
\ee
Since the \lhs of (\ref{49}) is equal to   (\ref{412}), we have now
\be
\hJ^\mu=\hW_P^\mu+\di_\nu\hU^{[\mn]}~~{\rm in~ which}~~\hW_P^\mu=(-2*\hLL^\mu_\nu -D_\la\hLL_\nu^{~\mu\la})\xi^\nu+\hLL_\la^{~\mn}D_\nu\xi^\la 
\lb{415}
\ee
This expression is obviously the same as (\ref{320}) in affine gravity since, if we take into account the identity (\ref{218}),
\be
\hW_P^\mu=\hW^\mu=\hLL^{~\mu}_\la\xi^\la+\hLL_\la^{~\mn}D_\nu\xi^\la, 
\lb{416}
\ee
 which is indeed (\ref{320}).

The passage from affine gravity  to the Palatini formalism  is  implicit in the work of Julia and Silva \cite{JS00}. One    difference is that  the whole layout    is here explicit and in tensorial notations.
%---------------------------------5 The superpotential equation-------------------------
\sect{Superpotentials  and   charges in Lovelock gravity theories}
\vs
{\it (i) Hamiltonian covariant formalism}
\vs
With an expression for the  current conserved on-shell (\ref{415}) we may write the covariant Hamiltonian $H$. We   keep the letter $H$ for in the end we are mostly interested in the total mass energy. However,   at this stage we shall not specialize $\bxi$ to an asymptotic timelike  Killing vector. Thus, in the Palatini formalism,
\be
H=\int_V\hW^\mu dV_\mu +\oint_S\hU^{\mn}dS_{\mn}\!=\!\int_V\lll\hLL^{~\mu}_\la\xi^\la+\hLL_\la^{~\mn}D_\nu\xi^\la\rrr dV_\mu +\oint_S\hU^{\mn}dS_{\mn}.
\lb{51}
\ee
We can equally use the field components of affine gravity, before  breaking the gauge  invariance. The calculation is easily seen to lead to the same result once gauge invariance is broken. 
From (\ref{36}) we know that  $\hLL^{~\mu}_\la$ is an ordinary derivative which depends on 
$\bg$ and $\bB$ only. 
On the other hand, in Lovelock gravity, $\hLL_\la^{~\mn}$ are variational derivatives, see (\ref{25}), which contain  not only $\bg$ and $\bB$ but are linear and homogeneous in derivatives  of the metric components $D_\la g^{\mn}$ but contain no derivatives of $\bB$.
The covariant Hamiltonian \eeee are obtained from the variation of $H$ which reduces thus to:
\be
\de H=\int_V\LLL \fr{\de\hW^\mu}{\de g^{\rs}}\de g^{\rs}+ \fr{\de\hW^\mu}{\de \Ga^{\la}_{~\rs}}\de \Ga^{\la}_{~\rs}\RRR dV_\mu+\oint_S\LLL  \fr{\di\hW^\mu}{\lll\di_\nu g^{\rs}\rrr}\de g^{\rs} +\fr{\di\hLL^{~\mu}_\et\xi^\et}{\di\lll\di_\nu \Ga^{\la}_{~\rs}\rrr}\de\Ga^{\la}_{~\rs}   + \de\hU^{\mn}                    \RRR dS_{\mn}
\lb{52}
\ee
It is not at all obvious  that the term in front of $  \de\hU^{\mn}  $ in the brackets on the right hand side is antisymmetrical in $\ts{\mn}$, however it is.  The prove goes as follows: consider the left hand side of (\ref{34}) which is a divergence. Its variational derivative with respect to any field component,  say $y^A$, is thus equal to zero. This holds for any $\bxi$. Replace $\bxi$ by $\ep(x)\bxi$. It holds also for any $\ep(x)$. The factor of $D_{\mn}\ep$ must thus be equal to zero. This factor is $\di \hW^{(\mu}/\di[\di_{\nu)} y^A]$. Thus the factors in the bracket on the right hand side are anti-symmetric in $\ts{\mn}$.

\vs
{\it (ii) Dirichlet boundary conditions and the \sup}
\vs
Following Julia and Silva the functional equation for the  superpotential is given by equating the second bracket to zero:
\be
\fr{\di\hW^\mu}{\lll\di_\nu g^{\rs}\rrr}\de g^{\rs} +\fr{\di\hLL^{~\mu}_\et\xi^\et}{\di\lll\di_\nu \Ga^{\la}_{~\rs}\rrr}\de\Ga^{\la}_{~\rs}   + \de\hU^{\mn}=0.   
\lb{53}  
\ee
Superpotentials must satisfy all boundary conditions. The most common conditions in Einstein's theory of gravity are the Dirichlet boundary conditions $g^{\rs}|_S=\Bar g^{\rs}$, or   equivalently  $\de g^{\rs}|_S=0$. We shall also identify the $ {\bGa}$'s on the boundary with Christoffel symbols which is equivalent to taking $ D_\la  g^{\rs}|_S=0$ or $\de (D_\la  g^{\rs})|_S=0$. As a consequence on the boundary $B^{\la}_{~\si\mn}|_S={\Bar R^{\la}_{~\si\mn}}$, the Riemann tensor of the background, and $\de B^\la_{~\si\mn}|_S=0$. Finally the field equations themselves must remain satisfied. Thus, for isolated sources the boundary conditions are:
\be
\de g^{\rs}|_S=\de (D_\la  g^{\rs})|_S=\de B^\la_{~\si\mn}|_S=\de\hLL_\la^{~\mu}=\de\hLL_\la^{~\mn}=0.
\lb{54}
\ee
Equations (\ref{53}) reduce thus to 
\be
\fr{\di(\hLL^{~\mu}_\eta\xi^\eta)}{\di\lll\di_\nu\Ga^{\la}_{~\rs}\rrr}\de\Ga^{\la}_{~\rs} + \de\hU^{\mn}=0.
\lb{55}
\ee
The factors of $\de\Ga^\la_{~\rs}$ in  (\ref{55})  are  entirely defined on the boundary. As far as the differential equation for $\bU\!\!$ is concerned, we may thus  integrate \eee (\ref{55}), and obtain an expression linear in $\Ga^\la_{~\rs}$. The ``constant of integration" will  be {\it minus}   the same expression with $\Ga^\la_{~\rs}$ replaced by  $\Bar\Ga^\la_{~\rs}$  for reasons explained in the Introduction. The   formula  for the \sup is thus  
\be
\hU^{\mn}= - \fr{\di(\hLL^{~\mu}_\eta\xi^\eta)}{\di\lll\di_\nu\Ga^{\la}_{~\rs}\rrr}\De^{\la}_{~\rs} ~~~ {\rm with}~~~\De^{\la}_{~\rs}=\Ga^{\la}_{~\rs}- \Bar\Ga^{\la}_{~\rs}.
\lb{56}
\ee
    Finally, since $\hLL$ is a function of $\bg$ and $\bB$, we may also rewrite (\ref{55}) like this:
 \be
 \hU^{\mn}=2\fr{\di(\hLL^{~\mu}_\eta\xi^\eta)}{\di B^\la_{~\rs\nu}}\De^\la_{~\rs}.
 \lb{57}
 \ee
 This   formula for the covariant \sup is a   blend of two derivatives: one with respect to $\bte$ at $\bga$ and $\bom$ constant in affine gravity giving the generator $\hLL^{~\mu}_\eta\xi^\et$ and one with respect to $B^\la_{~\rs\nu}$ at $\bg$ constant in the Palatini representation. As far as we know this result is new, simple and compact. The corresponding charge\footnote{The \sup given in \cite{KL08} is not the same as the one given in (\ref{57}). The reason is that we used an equality (3.20) (in \cite{KL08}) which does not correspond to a covariant generalization of the Regge-Teitelboim procedure. It is rather equivalent to what amounts in Julia and Silva \cite{JS00} to take $X\ne0$ in their equation (10). The \sup we found in \cite{KL08} gives however correctly the charges for backgrounds that are maximally symmetrical which is the most common case and which explain that all the applications described in \cite{KL08} are correct. The claim of the paper that it generalizes the Regge Teitelboim method along the line described by Julia and Silva is, however, wrong.} $Q$ is
 \be
 Q=\oint_S \hU^{\mn}dS_{\mn}
 \lb{58}
 \ee
 in which $ \hU^{\mn}$ is defined in (\ref{57}).
 
  %-------------------------- 6 Examples of charges----------------------------
  
\sect{Some examples }
% {\it (i) Charges for   Einstein's theory of gravitation in   N dimensions}
 {\it (i) Charges in  N dimensional  Einstein's theory of gravitation    }
\vs
This simple example will show how the machinery works.  The Lagrangian is
\be
\hLL_1=\fr{1}{2\ka}\hat B=\fr{1}{2\ka}\hg^{\mn}B_{\mn}= \fr{1}{4\ka}(\de^\rh_\la \hg^{\si\nu} - \de^\si_\la \hg^{\rh\nu})B^\la_{~\nu\rs}=\fr{1}{2\ka}\de^\rh_\la 
\hg^{\si\nu}B^\la_{~\nu[\rs]}.
\lb{61}
\ee
 In four dimensions we take as usual for $\ka$ Einstein's coupling constant $\ka=\fr{8\pi G}{c^4}$. In $N$ dimensions 
 \be
 \ka=\fr{2S_{N-2}G_N}{c^4},
 \lb{62}
 \ee
  where $S_{N-2}$ is the surface of a unit sphere of dimension $(N-2)$ and $G_N$ is a gravity coupling constant.
In affine gravity, the field components are $\ga^{ab}$, $\te^a_\mu$ and $\om^a_{~\mu b}$ in terms of which,   see (\ref{26}) and   (\ref{29}):
\be
\hLL_1=\fr{1}{2\ka} |\te|\te^{~\mu}_a\te^{~\nu}_b\lll    \sq{-\ga} B^{[ab]}_{~[\mn]} \rrr~{\rm where}~\te=\det(\te^a_{~\mu})~,~\ga=\det({\ga_{ab}})~{\rm and} ~B^{ab}_{~\mn} =\ga^{bc}B^a_{~c\mn}.
\lb{63}
\ee
The term in parenthesis is independent of the $\bte$'s.
The calculation of the derivative with respect to $\te^a_{~\mu}$ is thus very simple. One has, however, to be careful about symmetrizations and anti-symmetrizations  of   indices. The quantity that interests us, the generator appearing in (\ref{56}), is here $_1\hLL^{~\mu}_\eta \xi^\eta$ which in a condensed form may be written as
\be
_1\hLL^{~\mu}_\eta\xi^\eta=\fr{1}{2\ka}\lll  \xi^\mu\de^{[\si}_\la \hg^{\nu]\rh} +  \xi^\si\de^{[\nu}_\la \hg^{\mu]\rh} + \xi^\nu\de^{[\mu}_\la \hg^{\si]\rh}            \rrr B^\la_{~\rs\nu}.
\lb{64}
\ee
From this follows that the \sup $\bf U_1$ is
\be
\hU_1^{\mn}=\fr{1}{2\ka}\lll   \xi^{[\mu} \hg^{\nu]\rh}\De^\si_{~\rs} + \xi^{[\nu}\hg^{\si\rh}\De^{\mu]}_{~\rs}+\xi^\si \hg^{\rh[\mu }\De^{\nu]} _{~\rs}       \rrr\!.
\lb{65}
\ee
Notice the circular symmetry of indices $\ts{\mu, \si, \nu}$ in (\ref{64}) and $\ts{\mu, \nu, \si}$ in (\ref{65}).
In exactly this form, but in Minkowski coordinates on a flat \bbb and for $\xi^\mu=\{1, 0, 0, 0\}$, the \sup for the total mass energy in Einstein's theory in 4 dimensions   was published  by von Freud \cite{Vo} in 1939.

A more standard form is obtained by noticing that
\be
\xi^\si \hg^{\rh[\mu}\De^{\nu]} _{~\rs}    = D^{[\mu}\hat\xi^{\nu]} - \Bar{ D^{[\mu}\hat\xi^{\nu]} }
\lb{66}
\ee
and then writing $\hU^{\mn}_1$ like this:
\be
\hU_1^{\mn}=\fr{1}{2\ka}\lll D^{[\mu}\hat\xi^{\nu]} - \Bar{ D^{[\mu}\hat\xi^{\nu]} }\rrr +\fr{1}{2\ka}\xi^{[\mu}\hat k^{\nu]} ~~~{\rm with}~~~\hat k^{\nu}=\hg^{\nu\rh}\De^\si_{~\rs} - \hg^{\rs}\De^\nu_{~\rs}.
\lb{67}
\ee
In this form we recognize the KBL \cite{KBL} superpotential\footnote{On a  flat background this expression was   given in \cite{Ka}.}. The first term $\fr{1}{2\ka} D^{[\mu}\hat\xi^{\nu]}$ is half  the Komar \cite{Ko} superpotential. For applications formula (\ref{65}) is simpler. Various physical properties including standard results for charges derived from the KBL \sup   have been reviewed in \cite{JS00}. It gives among other things the Bondi mass \cite{KL97} for radiating systems in four dimensions. The KBL \sup is thus valid at both spatial and null infinity and on any background that satisfies the boundary conditions. Here is a new application of the superpotential at null infinity in five dimensions. We have calculated the   mass loss using  the asymptotic structure of the solution  of Einstein's \eeee given by Tanabe, Tanahashi and Shiromizu \cite{TTS10}, see also \cite{TTS11}, in ``Bondi coordinates" and obtained   their formula for the mass loss\footnote{Adjusted to our coupling constant in front of the Lagrangian and our units.}. Some details of the calculations are given in Appendix A.
% {\it (ii) Gauss-Bonnet Gravity in $D$ dimensions}
\vs
{\it (ii) Superpotential of Gauss-Bonnet Gravity in $N>4$ dimensions}
\vs
Einstein-Gauss-Bonnet theories of gravity are usually considered in combination with Einstein's Lagrangian [see sub-section (iii) below]. Since we obtained already $\bf \hU_1$ we shall consider $\bf \hU_2$ derived from the Lagrangian density $\hLL_2$ which we shall write in a  form similar to (\ref{63}) the most   convenient to calculate the generator $_2\hLL^{~\mu}_\eta\xi^\eta$ and which we took from Lovelock's paper of 1971 \cite{Lo}:
\be
\hLL_2=\fr{1}{2^2\ka}|\te|\te_a^{~\mu}\te^{~\nu}_b \te^{~\rh}_c\te^{~\si}_d 
\lll\sq{-\ga}B^{[ab]}_{~~~[\mn]}B^{[cd]}_{~~~[\rs]}\rrr\!.
\lb{68}
\ee
The terms within parentheses depends on $\bga$'s and $\bom\!\!\!$'s and not on $\bte$. The $\bte$'s are all in front of the parenthesis.
Equation (\ref{68}) provides   the generator $_2\hLL_\eta^{~\mu}\xi^\et$ through  ordinary partial  derivatives of  (\ref{68}) with respect to the $\bte$'s. The result is worth rewriting in terms of $\bf g$'s and $\bf B$'s, since the superpotential $\bf \hU_2$ is also obtained by ordinary  partial derivatives  of the generator $_2\hLL_\eta^{~\mu}\xi^\et$ with respect $\bf B$'s at fixed $\bf g$'s:
\be
_2\hLL^\mu_\eta\xi^\eta=\hat\Te^{\mu\eta\tau\mu'\nu'\rh'\si'}_{\la\ze}B^\la_{~\eta\mu'\nu'}B^\ze_{~\tau\rh'\si'}
\lb{69}
\ee
in which
\be
\hat\Te^{\mu\eta\tau\mu'\nu'\rh'\si'}_{\la\ze}\!\!=\!\fr{\sq{-g}}{2^2\ka}\LLL \!\lll \xi^\mu\de^{[\mu'}_\la \!g^{\nu']\eta}          \! + \!  \xi^{\mu'}\de^{[\nu'}_\la \!g^{\mu]\eta}   \! + \!\xi^\nu\de^{[\mu} \!g^{\mu']\eta}\rrr \!   \de^{[\rh'}_\ze \!g^{\si']\tau}\! \! + \!\!\de_\la^{[\mu'}\!g^{\nu']\eta}\!\lll \!  \xi^{\rh'}\!\de_\ze^{[\si'} \!g^{\mu]\tau} \! - \!\xi^{\si'} \!\de_\ze^{[\rh'} \!g^{\mu]\tau}                       \!        \rrr \RRR\!.
\lb{610}
\ee
 
The resulting $\bf \hU_2$, see (\ref{57}), contains now  9 triplets with circular symmetry,    altogether 27 terms linear in the curvature tensor:
\ba
&&  \ka \hU^{\mn}_2= 3\LLL\hat\xi^{(\rh}R^{\mn)\si}_{~~~~\la}\De^\la_{~\rs}+\lll g^{\rh(\mu}G^\nu_\eta\De^{\si)}_{~\rs}-g^{\rh(\nu}G^\mu_\eta\De^{\si)}_{~\rs}\rrr\hxi^\eta\RRR \nn\\
 &&~~~~~~~~+ 3\lll g^{\rh(\si}R^{\mn)}_{~~~\la\eta}\hxi^\eta+R^{(\mu}_\la\hxi^\rh g^{\nu)\si} - R^{(\nu}_\la\hxi^\rh g^{\mu)\si}\rrr\De^\la_{~\rs} \nn\\
&&~~~~~~~~- 3 \lll\De^{(\rh}_{~~\rs}R^{\mn)\si}_{~~~~\eta}\hxi^\eta+  R^{\rh(\mu}\De^\nu_{~\rs}\hxi^{\si)}  -  R^{\rh(\nu}\De^\mu_{~\rs}\hxi^{\si)}\rrr\!. 
\lb{611}
 \ea
 The parenthesis represent circular permutations as   in (\ref{46}), which explains all those 3's.
If we separate $\hG^\si_\eta$ into Ricci and scalar curvature parts there are 11 instead of 9 triplets and altogether 33 terms.  An application of this complicated   formula is given in the next subsection.  
 ~~
% (iii) Einstein-Gauss-Bonnet theories of gravity with maximally symmetric boundaries}
\vs
{\it (iii) Einstein-Gauss-Bonnet theories of gravity with maximally symmetric boundaries}
\vs
In these theories of gravitation, the Lagrangian density is a linear combination of the Einstein-Hilbert Lagrangian $\hLL_1$ (a Lovelock Lagrangian of order 1) and the Gauss-Bonnet Lagrangian $\hLL_2$ (a Lovelock Lagrangian of order 2)
\be
\hLL_{2'}=\al_1\hLL_1+\al_2\hLL_2,
\lb{612}
\ee
in which $\al_1$ and $\al_2$ are self-coupling constants to be found experimentally. The corresponding \sup is thus 
\be
\hU^{\mn}_{2'}=\al_1\hU^{\mn}_1+\al_2\hU^{\mn}_2.
\lb{613}
\ee

In past applications \bbbb have mostly been flat or anti-de Sitter, i.e.$\!$ maximally symmetric spacetimes. In such \ssss \!:
\be
R^\la_{~\nu\rs}=    \fr{2R}{N(N-1)} \de_{[\rh}^\la \Bar g_{\si]\nu} ~~~ {\rm and } ~~~R_{\mn}=\fr{R}{N}\Bar g_{\mn}.  
\lb{614}
\ee
This makes $\bf \hU_2$ considerably simpler:
\be
\hU^{\mn}_2=\fr{(N-3)(N-4)}{N(N-1)}R\hU^{\mn}_1.
\lb{615}
\ee
We have applied this result to calculate the mass of the  Einstein-Gauss-Bonnet black holes  found by Deser and Tekin \cite{DT03} with a background curvature
\be
R_{\mn\rs}=\fr{1}{l^2}\lll      \Bg_{\mu\rh}  \Bg_{\nu\si}   -  \Bg_{\mu\si}  \Bg_{\nu\rh}      \rrr ~~~{\rm where}~~~l^2= - \fr{(N-1)(N-2)}{2\La}~~,~~{\La<0}.
\lb{616}
\ee
in the case considered in  \cite{DT03}, $\al_1=1$ and
\be
\al_2=\fr{-2l^2}{(N-3)(N-4)},
\lb{617}
\ee
one obtains the surprising result that 
\be
\hU^{\mn}_{2'}= - \hU^{\mn}_1\,!
\lb{618}
\ee
But this is indeed the correct result and it gives  the positive mass  found in Deser and Tekin by a quite different but     reliable method.
%(iv) Lovelock Gravity in any number of dimensions and in infinite dimensions}
\vs
{\it (iv) Lovelock Gravity in any number of dimensions}
\vs
  The Lovelock Lagrangian density of order $n$ in $N$ dimensions is as follows:
\be
\hLL_n=\fr{\sq{-g}}{2^n\ka}\!\!\!\sum_{n=0}^{~~n=p-1}\de^{\nu_1\nu_2\cd\cd\cd\nu_{2p}}_{\mu_1\mu_2\cd\cd\cd\mu_{2p}} B^{\mu_1\mu_2}_{~~~~~\nu_1\nu_2}\cd\cd\cd B^{\mu_{2p-1}\mu_{2p}}_{~~~~~~~~~~\nu_{2p-1}\nu_{2p}}
\lb{619}
\ee
with
\be1\le n \le p-1~ {\rm where}~p=\tha \!\!N ~({\rm for}~N ~{\rm even })~,  ~p=\tha\!\!(N+1) ~({\rm for}~N ~{\rm odd })~~{\rm and}~~\de^{\nu_1\nu_2\cd\cd\cd\nu_{2p}}_{\mu_1\mu_2\cd\cd\cd\mu_{2p}} ={\rm \det}(\bde). 
\lb{620}
\ee
 In 1985 Zwiebach \cite{Zw} speculated that the Lagrangian that would ultimately describe gravitational interactions and more fundamentally string interactions should have an infinite number of terms. Each ${\bf\hU}_n$ would give a finite contribution. In particular for \ssss that are asymptotically maximally symmetric,
\be
\hU^{\mn}_n=n\,\fr{(N-2n)[N-(2n-1)]\cd\cd\cd(N-4)(N-3)}{[N(N-1)]^{n-1}}R^{n-1}\hU_1^{\mn}.
\lb{620}
\ee 
 and for $N\ra\in$, the factor containing $N$ is equal to one. Thus, if the Lagrangian density
 \be
 \hLL_{\in}=\sum_{n=0}^{\in}\al_n\hLL_n,
 \lb{621}
 \ee
 the \sup is
 \be
 \hU^{\mn}_{\in}=\lll \!\sum_{n=0}^{\in}n\al_n R^n \!\rrr \hU^{\mn}_1.
 \lb{622}
 \ee
 This sum may converge for suitable $ \bal$'s.
 %---------------------7 Some comments--------------------------------
\sect{Some comments}

\nnn a] Regarding $\hU^{\rs}_1$ as given by \eee (\ref{65}). On a flat background, in Minkowski \coo  and  with \bm $\xi$ \ubm the Killing vectors  of spacetime translations,  $\hU^{\rs}_1$, as we observed, is exactly the \sup that Freud \cite{Vo} found, more than 70 years ago, to calculate mass-energy and total linear momentum in Einstein's theory. One wonders why Freud did not  calculate the angular momentum on the same occasion.

\nnn b] We emphasized several times that the superpotential holds  at null as well as at spatial infinity and that  $\hU^{\rs}_1$ at null infinity gives the Bondi mass loss in  four dimensions \cite{KL97} as well as in 5 dimensions as we show in Appendix A. We are not aware of a published work on radiating fields in Einstein-Gauss-Bonnet theories.  $\hU^{\rs}_{2'}$ is, of course, the \sup appropriate for calculating the Bondi mass loss in that case.

\nnn c] We imposed Dirichlet boundary conditions and found the KBL \sup in N dimensional Einstein's gravity. Julia and Silva showed that by imposing Neumann boundary conditions  we are led   to Komar's superpotential\footnote{A generalization of Komar's superpotential, that does not need a background, to any type of boundary conditions,  can be found in several papers of Obukhov and Rubilar's ; see for instance \cite{OR}. The role of \bbbb is here replaced by another ingredient, ``generalized" Lie derivatives.}.   Neumann boundary conditions have been considered in recent works. See for instance a paper by Kofinas and Olea \cite{KO} on Lovelock anti-de Sitter gravity.   We presume that  imposing Neumann boundary conditions  would have   led to Noether charges as considered in Iyer and Wald \cite{IW}.

\nnn d]   Jacobson and Myers \cite{JM} used the first law of  thermodynamics to define    mass-energy of Lovelock black holes. Mass-energy has, however, little to do with thermodynamics because     asymptotic spacetimes ignore the source of gravity.  A direct calculation of the mass-energy of Lovelock black-holes, independent of thermodynamic considerations, is  naturally provided by the  \sup $\hU^{\rs}$. An   approach, similar to that of Jacobson and Myers, was used by Gibbons {\it et al} \cite{GLPP} to find the mass of Kerr-anti-de Sitter black holes. Needless to say the same mass-energy is obtained  with the KBL superpotential $\hU^{\rs}_1\!$, see \cite{DK}. 

\nnn e] We dealt with Lovelock Lagrangians whose variational derivatives have no  derivatives of  the curvature tensor. Conservation laws have, however, been considered for non-Lovelock Lagrangians by Deser and Tenkin \cite{DT03} at spatial infinity and by Wald and Zoupas \cite{WZ00} at null infinity. A covariant generalization of Regge and Teitelboim principle to a Palatini  or affine gravity formulation needs yet to be worked out but do not appear to be trivial unless one introduces auxiliary fields in order to reduce second order \eeee to first order one which is always possible while preserving the gauge symmetries, see   \cite{JS99}.

\vs
{\bf  Acknowledgments}
\vs
G. I. L. is very grateful to Tanabe for his   advice in calculating the Bondi mass.
\vs
%----------------------------------Appendix-----------------------------------
\cl{\LARGE{{\bf Appendix}}}
\begin{appendix}
\setcounter{equation}{0}
%------------------------Appendices A  Integrals over products of Legendre functions ----------------------------------
\section{ The Bondi loss of mass    in 5 dimensional Einstein gravity }
\renewcommand{\theequation}{\Alph{section}.\arabic{equation}}
\normalsize
\setlength{\baselineskip}{20pt plus2pt}
Tanabe, Tanahishi and Shiromizu \cite{TTS10} calculated the asymptotic solution of Einstein's \eeee in five dimensions at null infinity. They used coordinates similar to those of Bondi and calculated, among other things, the Bondi loss of mass-energy. We have calculated the mass loss from our formula  
(\ref{64}) for $\bf \hU_1$ and have  found, of course, the same result. Here we use their notations up to minor changes with their signature of the metric and give only the   quantities needed for the calculation.
The Bondi coordinates are $x^\mu=\{ u, r, x^A\}$ thus with $x^A=\{\te, \vf, \psi\}$, $A,B,\cdots=2, 3, 4$ which are the coordinates of a unit sphere of dimension 3, $S_3$.
 In these coordinates, the line element may be written
 \be
 ds^2=\lll -\fr{Ve^B}{r^2}+r^2h_{AB}U^AU^B\rrr du^2 - 2e^Bdudr+r^2h_{AB}(dx^A+U^Adu)(dx^B+U^Bdu).
 \lb{A1}
 \ee
$V, B, U^A$   depend on all coordinates. The asymptotic expansion in powers of $r$
 is as follows:
 \be
V=r^2+V_1r^{1/2}-\EE+\OO(r^{-1/2})~~,~~B=\fr{B_1}{r^3} +\OO(r^{-4})~~,~~U^A=\fr{U^A_1}{r^{5/2}} +\OO(r^{-3}),
\lb{A2}
 \ee
  $V_1, B_1, U^A_1$ and $h_{AB}$  are functions of $\{u,x^A\}$ only:
\ba
&&V_1={\ts{ - \fr{2}{3}}}\LLL   \fr{1}{\sin 2\te} \di_\te\lll \sin 2\te \,U^\te_1  \rrr +\di_\vf U_1^\vf +\di_\psi U_1^\psi        \RRR,\nn
\\
&&B_1={\ts{ - \fr{1}{8}}}\lll C_{11}^2+C_{11}C_{21}+C_{12}^2+D_{11}^2+D_{12}^2+D_{13}^2\rrr,\nn
\\
&& U_1^\te={\ts\fr{2}{5}}\LLL   \fr{(C_{11}-C_{12})}{\tan\te}   -(2C_{11}+C_{12}) \tan\te+\di_\te C_{11} +\fr{\di_\vf D_{11}}{\sin\te} +\fr{\di_\psi D_{12}}{\cos\te} \RRR\!,\nn
\\
&&U_1^\vf\sin\te={\ts\fr{2}{5}}\LLL    D_{11}\lll   \fr{2}{\tan\te}  - \tan\te \rrr     +\di_\te D_{11}   +\fr{\di_\vf C_{12}}{\sin\te}  + \fr{\di_\psi D_{13}}{\cos\te}   \RRR\!,\nn
\\
&&U_1^\psi\cos\te={\ts\fr{2}{5}}\LLL   D_{12}\lll   \fr{1}{\tan\te} - 2\tan\te \rrr  +\di_\te D_{12}    +\fr{\di_\vf D_{13}}{\sin\te}  - \fr{\di_\psi(C_{11}+C_{12})}{\cos\te}             \RRR\!,\nn
\\
&& h_{AB}=\Bar h_{AB}+\fr{q_{AB}}{r^{3/2}}+\OO(r^{-5/2})~~{\rm with}~~\Bar h_{AB}={\rm diag}\{1, \sin^2\te, \cos^2\te\},
\lb{A3}
\ea
and
\ba
 && q_{\te\te}=C_{11} ~~~,~~~q_{\te\vf}=\sin\te\,D_{11}~~~,~~~q_{\te\psi}=\cos\te \,D_{12} ,\nn
 \\
 &&q_{\vf\vf}=\sin^2\te \,C_{12}~~~,~~~q_{\vf\psi}=\sin\te\cos \te \,D_{13}~~~,~~~q_{\psi\psi}=\cos^2\te \,C_{13};
 \lb{A4}
\ea
The three $\bf C$'s and three $\bf D$'s are functions of $\{u, x^A\}$, they  are not independent:
\be
C_{11}+C_{21}+C_{31}=0~,~C_{12}+C_{22}+C_{32}=0~,~C_{13}+C_{23}+C_{33}=D^2_{11}+D^2_{21}+D^2_{31}.
 \lb{A5}
\ee
There are thus 5 independent functions  associated with the independent degrees of freedom of the gravitational waves in five dimensions.
 $\EE(u,x^A)$ is an integration ``constant".

The background at infinity is flat. Its line element in our coordinates is
\be
\Bar{ds}^2=\Bg_{\mn}dx^\mu dx^\nu= - du^2 - 2dudr+r^2\lll   d\te^2+\sin^2\te    d\vf^2+\cos^2\te d\psi^2 \rrr~~\Ra~~\sq{- \Bg} =  \tha\!r^3\sin 2\te .
\lb{A6}
\ee

To calculate the mass loss we shall need the inverse components of the metric $g^{\mn}$; the non-zero ones are as follows:
\be
g^{01}\!=\! - 1\!+\!\fr{B_1}{r^3}\!+\!\OO(r^{-7/2})~,~g^{11}\!=\! 1\!+\!\fr{V_1}{r^{3/2}} \!-\! \fr{\EE}{r^2} \!-\! \fr{B_1}{r^3}\!+\!\OO(r^{-7/2})~,~g^{1A}\!=\! \fr{U_1^A}{r^{5/2}}\!+\!\! \fr{U_2^A}{r^{7/2}}\!\!+\!\OO(r^{-11/2}),\nn
\ee
\be
 g^{22}\!=\! \fr{1}{r^2}\lll  1\!-\! \fr{C_{11}}{r^{3/2}} \!-\! \fr{C_{12}}{r^{5/2}}\!\rrr+\fr{1}{r^5}\lll   \tha\!C_{11}^2 - C_{13} +D_{11}^2+D_{12}^2         \rrr+\OO(r^{-11/2}),~
 ~~~~~~~~~\nn
 \ee
 \be
  g^{23}\!=\! -\fr{1}{r^3\sin\te}\lll  \fr{D_{11}}{r^{1/2}}\!+\!   \fr{D_{12}}{r^{3/2}} \rrr\! +  \fr{1}{r^{5}\sin\te} \LLL          - C_{13}D_{11} +(D_{12}-1)  D_{13}     \RRR\!
   +\OO(r^{-11/2}),\nn
\ee
\be
g^{24}\!=\! -\fr{1}{r^3\sin\te}\lll  \fr{D_{12}}{r^{1/2}}\!+\!   \fr{D_{22}}{r^{3/2}} \rrr\!+\fr{1}{r^5\sin\te}\lll   - C_{12}D_{12} - D_{23}    +D_{11}D_{13}  \rrr+\OO(r^{-11/2}),\nn
\ee
\be
g^{33}\!=\!\fr{1}{r^2\sin^2\te}\lll 1\!-\! \fr{C_{12}}{r^{3/2}}  \!-\!  \fr{C_{22}}{r^{5/2}} \rrr+\fr{1}{r^5\sin^2\te}\lll   \tha\!C_{12}^2 - C_{23} +D_{11}^2+D_{13}^2         \rrr +\OO(r^{-11/2}) \nn
\ee
\be
g^{34}\!=\! \!-\fr{2}{r^2\sin2\te}\lll  \fr{D_{13}}{r^{3/2}} \!+\!  \fr{D_{23}}{r^{5/2}}  \rrr \! + -\fr{2}{r^5\sin2\te} \lll     C_{11}D_{13} +D_{33} - D_{11}D_{12}     \rrr    +\OO(r^{-11/2}),\nn
\ee
\be
g^{44}\!=\!\fr{1}{r^2\sin^2\te}\lll 1\!-\!  \fr{C_{13}}{r^{3/2}}  \!-\!   \fr{C_{23}}{r^{5/2}}  \rrr\!+\fr{1}{r^5\sin^2\te}\lll   \tha\!C_{13}^2 - C_{33} +D_{12}^2+D_{13}^2         \rrr +\OO(r^{-11/2}). 
\lb{A7}
\ee
 $g^{\mn}$ appears  in the Christoffel symbols $\Ga^\la_{~\mn}$ and  in the end we need $\De^\la_{~\mn}=\Ga^\la_{~\mn}-\Bar\Ga^\la_{~\mn}$. The background calculations are not particularly difficult. One will not need all the $\bf\De$'s for the energy associated with the  time translation Killing vector $\bxi=\{1, 0, 0, 0, 0\}$. Moreover $\sq{- \Bar g}\propto r^3$ and most of the $\sq{-\Bar g}\bf U_1$'s go to zero when $r\ra\in$; in the end the only remaining term in $\bf\hU_1$ that contributes to the energy  is 
\be
\sq{-\Bar g}\De^1_{~01}=\tha \!\!\sin 2\te\, \EE + {\rm non ~contributing~ parts}.
\lb{A8}
\ee
From this follows that the charge
\be
E=\oint_S\EE dS_3
\lb{A9}
\ee
Finally from Einstein's \eeee $R_{uu}=0$ and $R_{ux^A}=0$ we obtain, like Tanabe, Tanahashi and Shiromizu in \cite{TTS10} the Bondi mass loss; with $E=Mc^2$
\be
\fr{dM}{du}= - \fr{c^2}{G_5}\fr{1}{S_3}\oint _S   \Big{\{} {\ts{\fr{1}{3}}}\LLL (\di_uC_{11}\!+\!\tha\di_uC_{21})^2\!+\!{\ts{\fr{3}{4}} }( \di_uC_{21})^2+(\di_uD_{11})^2\!+\!(\di_uD_{12})^2\!+\!(\di_uD_{13})^2\RRR\Big{\}} dS_3 .
\lb{A10}
\ee
\end{appendix}

\end{document}